\pdfoutput=1

\documentclass[prl,twocolumn,showpacs,preprintnumbers,amsmath,amssymb,
superscriptaddress,nofootinbib]{revtex4-1}
\usepackage{epsfig}
\usepackage{graphicx}
\usepackage{color}

\begin{document}

\preprint{CALT-TH-2014-167}

\title{Effective Field Theories from Soft Limits}

\author{Clifford Cheung}
\affiliation{Walter Burke Institute for Theoretical Physics, California 
Institute of Technology, Pasadena, CA 91125}

\author{Karol Kampf}
\affiliation{Institute of Particle and Nuclear Physics, Charles
University in Prague, Czech Republic}

\author{Jiri Novotny}
\affiliation{Institute of Particle and Nuclear Physics, Charles
University in Prague, Czech Republic}

\author{Jaroslav Trnka}
\affiliation{Walter Burke Institute for Theoretical Physics, California 
Institute of Technology, Pasadena, CA 91125}


\begin{abstract}
We derive scalar effective field theories---Lagrangians, symmetries, and all---from on-shell scattering amplitudes constructed purely from Lorentz invariance,  factorization, a fixed power counting order in derivatives, and a fixed order at which amplitudes vanish in the soft limit.   
These constraints leave free parameters in the amplitude which are the coupling constants of well-known theories: Nambu-Goldstone bosons, Dirac-Born-Infeld scalars, and Galileons.  Moreover, soft limits imply conditions on the Noether current which can then be inverted to derive Lagrangians for each theory.  We propose a natural classification of all scalar effective field theories according to two numbers which encode the derivative power counting and soft behavior of the corresponding amplitudes.  In those cases where there is no consistent amplitude, the corresponding theory does not exist.
\end{abstract}

\maketitle

\section{Introduction}

Infrared dynamics are inextricably linked to symmetry.  For example, soft limits in gauge and gravity theories are fixed by conservation laws \cite{Weinberg:1965nx}, while soft limits of pion amplitudes secretly encode underlying patterns of symmetry breaking \cite{ArkaniHamed:2008gz}.  Tacitly,  symmetries are considered primary and the corresponding soft theorems secondary.  In this letter we argue for precisely the opposite: by constructing  scattering amplitudes directly and imposing various soft behaviors, we instead derive the theories and their symmetries.  

The idea of building a theory from its scattering amplitudes rather than its Lagrangian is not new.  Famously, tree amplitudes in gauge and gravity theories can be constructed solely from considerations of Lorentz invariance and factorization.  The same is true of non-linear sigma models \cite{nlsm}, albeit with the crucial and additional assumption of the so-called Adler zero \cite{nlsmadler}, which describes the vanishing of pion scattering amplitudes in the soft limit.  

The present work is a generalization of this prescription with the aim of enumerating all possible effective field theories of a massless scalar. We focus here on on-shell tree amplitudes in four dimensions, but our methods apply to diverse dimensions and loop integrands.
In the soft limit of an external leg, $p\rightarrow 0$, the tree amplitude is
\begin{align}
A(p) & = {\cal O}(p^\sigma),
\label{eq:soft}
\end{align}
where $\sigma $ is a non-negative integer characterizing the soft limit degree. Larger values of $\sigma$ imply cancellations in the amplitude enforced by relations among the coupling constants of the underlying theory, {\it i.e.}~more symmetry.

A massless scalar has the schematic Lagrangian,
\begin{align}
{\cal L} &=(\partial\phi)^2\sum_{m,n=0}^\infty \lambda_{m,n} \partial^{m}  \phi^{n},
\label{eq:L}
\end{align}
where $m$ is even by Lorentz invariance.
In general, the soft limit will enforce cancellations among diagrams of different topologies.  For example,  an $n+2$ particle amplitude includes diagrams with a  single $\lambda_{m,n}$ vertex as well as diagrams with a single propagator connecting a $\lambda_{m', n'}$ vertex to a $\lambda_{m'', n''}$ vertex.  By dimensional analysis, cancellations can only occur if $m=m' + m'' $ and $n=n' + n''$, corresponding to all $\lambda_{m,n}$ for which
\begin{align}
\rho &= m/n,
\label{eq:linear}
\end{align}
for a fixed non-negative rational number $\rho$ characterizing a particular power counting order in derivatives.  For fixed $\rho$, the Eq.~\ref{eq:L} takes the schematic form 
\begin{equation}
{\cal L}_{(\rho)} = (\partial\phi)^2F(\partial^{m}\phi^{n}),
\end{equation}
for a general function $F$, where 
$m$ and $n$ are the smallest numbers satisfying Eq.~\ref{eq:linear}.  Some familiar examples are 
\begin{align}
{\cal L}_{(0)} =(\partial \phi)^2F(\phi), \quad 
{\cal L}_{(1)} = (\partial \phi)^2F(\partial \phi),
\label{eq:ex}
\end{align}
corresponding to theories of free fields and Nambu-Goldstone bosons, respectively.

For a theory ${\cal L}_{(\rho)}$ we can impose the soft limit in Eq.~\ref{eq:soft} to constrain $F$, yielding a new theory ${\cal L}_{(\rho,\sigma)}$.  Thus, all scalar effective theories can be classified by two numbers, $(\rho,\sigma)$, which specify the derivative power counting of a theory together with the degree of its soft limits.  With an explicit Lagrangian, it is straightforward to compute the scattering amplitude and its soft limit but a more interesting exercise is the reverse: assume a value of  $(\rho,\sigma)$ and derive the corresponding theory.  

To begin, we construct general ansatze for on-shell tree amplitudes consistent with Lorentz invariance and factorization but restricted to a particular $(\rho,\sigma)$ derivative power counting and soft limit.  This generates the span of all possible amplitudes describing massless scalars.  For many values of $(\rho,\sigma)$ there is no consistent scattering amplitude, so there is no corresponding theory.  Even if a consistent amplitudes exists, however, this may not be so interesting if the soft limit is obvious from counting the number of derivatives per field.  By this logic a soft limit $\sigma \leq (m+2)/(n+2)$ is automatic, so the interesting case is in the opposite regime,
\begin{align}
\sigma > \frac{\rho n+2}{n+2}, 
\label{eq:interesting}
\end{align}
after plugging into Eq.~\ref{eq:linear}.  Tab.~\ref{tab:theories} summarizes those theories which have enhanced soft limits which exceed the degree expected from naive derivative power counting.  Also listed are the number of physical parameters which define each theory.
\begin{table}[t]
\begin{tabular}{c|c|c}
$(\rho,\sigma)$ & Theory & \; \textrm{Parameters} \; \\ \hline
\; $(0,\infty)$ \; & \textrm{Free} & \textrm{0} \\ \hline
(1,2) & \; \textrm{Dirac-Born-Infeld} \; & 1 \\ \hline
(2,2) & \textrm{Galileon}$_{4,5}$ & 2 \\  \hline
(2,3) & \textrm{Galileon}$_4$ & 1
\end{tabular}
\caption{Theories with $(\rho, \sigma)$ derivative power counting and soft limit degree that have enhanced soft limits.}
\label{tab:theories}
\end{table}
Here Galileon$_{4,5}$ denotes the original Galileon theory in a basis where the three point interaction vertex has been removed by a field redefinition and Galileon$_4$ denotes Galileon$_{4,5}$ truncated to just the four point interaction.  For Galileon$_4$ we have strong evidence---up to twelve point amplitudes---for an intriguing ${\cal O}(z^3)$ enhanced soft limit.

Afterwards, we show that fixing $(\rho,\sigma)$ places constraints on the Noether currents which can be used to immediately derive the Lagrangians for Dirac-Born-Infeld (DBI) and Galileon.

\section{Amplitudes from Ansatze}

We now construct an ansatz for on-shell tree amplitudes fixed by Lorentz invariance, factorization, and a specified derivative power counting and soft limit degree, $(\rho, \sigma)$.   If such an ansatz exists, then the corresponding theory can exist.

The on-shell three point amplitude vanishes in any theory due to kinematics, so here we focus on the case where the leading non-zero on-shell amplitude is four point.  An analogous discussion applies for theories in which the leading non-zero amplitude is higher point.

\subsection{Definition of Ansatze}

Any  scalar $n$ point on-shell scattering amplitude can be written in terms of the kinematical invariants
\begin{equation}
s_{ij} = (p_i+p_j)^2 = 2(p_i\cdot p_j), \quad i,j \in \{1,\ldots,n\},
\end{equation}
which is a redundant basis.  First of all, by momentum conservation we can always eliminate all dependence on the momentum of particle $n$, so we can restrict to $s_{ij}$ where $i,j \neq n$.  Second, there is an additional constraint because particle $n$ is on-shell so $\sum_{i,j\neq n} s_{ij} =0$. Last of all, in four dimensions, five generic momenta are necessarily linearly dependent, leading to the so-called Gram-determinant relations.  Since these are non-linear constraints, a truly independent set of $s_{ij}$ is difficult to compute analytically.  Instead it is much simpler to use a redundant basis of kinematic invariants and mod out by the redundancy at the end of the calculation.   

The general ansatz for the $n+2$ point amplitude in a theory with derivative power counting $\rho=m/n$ is 
\begin{align}
A_{n+2} &= \sum_\alpha c^{(0)}_{\alpha} (s_{\alpha_1}\dots s_{\alpha_{m/2+1}}) + \sum_{\alpha,\beta} \frac{c^{(1)}_\alpha (s_{\alpha_1}\dots s_{\alpha_{m/2+2}})}{s_{\beta}}\nonumber\\
&\hspace{0.5cm}+ \sum_{\alpha,\beta}\frac{c^{(2)}_\alpha (s_{\alpha_1}\dots s_{\alpha_{m/2+3}})}{s_{\beta_{1}}s_{\beta_{2}}} + \dots,
\label{eq:ansatz}
\end{align}
where $\alpha$ labels pairs of external legs that enter into the numerator factors and $\beta$ labels factorization channels whose corresponding off-shell propagators are $s_\beta = \sum_{i,j\in\beta} s_{ij}$.
Symmetries of the corresponding Feynman diagrams relate many coefficients $c_{\alpha}^{(k)}$, and moreover the ansatz is kinematically redundant due to reasons mentioned above. 

\subsection{Definition of Soft Limit}

To define the soft limit, we send $p\rightarrow zp$ for one of the external legs and expand in powers of $z$,  
\begin{equation}
A_{n} = \sum_{s=0}^\infty  A_{n,s} z^s,
\end{equation}
where we assume that the soft limit is non-singular.
To enforce the soft limit $A_n = {\cal O}(z^\sigma)$ we must solve for the coefficients $c_\alpha^{(k)}$ of the ansatz in Eq.~\ref{eq:ansatz} subject to $A_{n,s }=0$ for $s<\sigma$.   As noted earlier,  the $s_{ij}$ satisfy complicated non-linear constraints, so these equations cannot be solved analytically.    However, we can study the system numerically by evaluating the ansatz many times at arbitrary kinematical points, yielding linear equations in $c_\alpha^{(k)}$ which are easy to solve. We then plug the solution back into the ansatz and the number of independent parameters remaining determines the number of physical theory parameters.

\subsection{Results}


The analysis of three and four point amplitudes is simple for any $\rho$.  The three point amplitude vanishes by kinematics, while the four point amplitude is non-trivial and has a soft limit fixed by the number of derivatives per field, independent of the explicit forms of operators. In particular, for generic $\rho$ we have $m+2=2 \rho+2$ derivatives and therefore the kinematical ansatz is
\begin{equation}
A_4 = \sum_{a_1a_2a_3} c_{a_1a_2a_3} (s_{12})^{a_1}(s_{23})^{a_2}(s_{31})^{a_3},
\end{equation}
for $a_1+a_2+a_3=\rho+1$. However, in the soft limit $s_{12},s_{23},s_{31}= {\cal O}(z)$ so regardless of the particular derivative structure of the amplitude, $A_4 = {\cal O}(z^{\rho+1})$.
No further cancellations are possible so we cannot obtain stronger behavior by relating parameters.  Let us now consider higher point amplitudes for various values of $\rho$.

\medskip
\noindent {\bf Case: $\rho=0$}
\medskip

The soft limit is ill-defined because the Lagrangian secretly describes a free field theory and the kinematical ansatz is zero.  This is manifest after a well-chosen field redefinition, $\phi \rightarrow \phi'(\phi)$, which takes ${\cal L}_{(0)} = (\partial \phi)^2 F(\phi) = \partial \phi' \partial \phi'$.   Hence, while off-shell Feynman diagrams are non-trivial, they all vanish on-shell.  

Note that this is not generally true if $\phi$ carries a flavor. Then the amplitudes for $\rho=0$ are non-trivial and one can show that there is a choice of Lagrangian coefficients  such that $A_{n,0}=0$ so the amplitude behaves like $A_n={\cal O}(z)$ in the soft limit for any $n$. This solution is the non-linear sigma model \cite{nlsmadler}.  Furthermore this is true  even if you consider only flavor-stripped amplitudes \cite{Kampf:2012fn}.

\medskip
\noindent {\bf Case: $0<\rho<1$}
\medskip

The soft amplitudes do not vanish, so $A_n= {\cal O}(1)$.  This can be derived by contradiction.  A vanishing soft limit requires that for each leg, $A_n\rightarrow 0$ when $p\rightarrow 0$.  Enforcing this on each leg sequentially and demanding a permutation invariant amplitude yields a unique ansatz,
\begin{equation}
A_n = p_1^{\mu_1}p_2^{\mu_2}\dots p_n^{\mu_n}\,L_{\mu_1\mu_2\dots \mu_n},
\end{equation}
where $L_{\mu_1\mu_2\dots \mu_n}$ is a completely symmetric tensor constructed from factors of momenta and the metric. This implies that the number of derivatives cannot be less than the number of fields, so $\rho \geq 1$.   This can be easily understood from the symmetry point of view: the theory must be derivatively coupled to have a vanishing soft-limit. 

\medskip
\noindent {\bf Case: $\rho=1$}
\medskip

The first non-trivial case for a single scalar is $\rho=1$ for which $m=n$ and we have one derivative per field. If we want to impose ${\cal O}(z)$ behavior in the soft-limit the theory must be necessary derivatively coupled, {\it i.e.}~the corresponding Lagrangian is ${\cal L}_{(1,1)}\sim \sum \lambda_{2n}\,(\partial\phi)^{2n}$. This simplifies the ansatz for the amplitude (all labels must appear in $s_{ij}$). For example, for $n=4$ and $n=6$ we get
\begin{align}
A_4 & = c_4\,(s_{12}s_{34}+s_{13}s_{24}+s_{14}s_{23})\\
A_6 & = 2c_4^2\left[\frac{\widetilde{s_{123}}\widetilde{s_{456}}}{s_{123}} + \dots\right]
+ c_6\,(s_{12}s_{34}s_{56}+ \dots),
\end{align}
where the ellipses denote the sum over all permutations and $s_{123}=s_{12}+s_{23}+s_{31}$, $\widetilde{s_{123}} = s_{12}s_{23}+s_{23}s_{31}+s_{31}s_{12}$. 

Now we impose an enhanced soft limit by demanding that $A_{n,1}=0$, so $A_n={\cal O}(z^2)$. The four-point case is trivial as $s_{12},s_{23},s_{31}\sim z$ in the soft-limit and ${\cal O}(z^2)$ is trivially satisfied and there is no condition on $c_4$. At six point this is a highly non-trivial constraint which is satisfied if we set $c_6=2c_4^2$. 

The same argument can be applied to each higher point amplitude, so $A_{8}={\cal O}(z^2)$ can be used to fix the new coupling coefficient $c_8$ in terms over lower order couplings.  Inducting to arbitrarily high point amplitudes, it is then obvious that the infinite system equations has at most one solution.  As it turns out there is indeed one solution, and the corresponding $c_{2n}$ conspire to yield
\begin{equation}
{\cal L}_{(1,2)} = -\frac{1}{g} \sqrt{1-g(\partial\phi)^2},
\end{equation}
where $g=2c_4$ and ignoring vacuum energy. This is also known as the scalar part of the DBI action which describes a fluctuation of a brane in extra dimension. The hidden symmetry is the non-linearly realized higher dimensional Lorentz symmetry on $\phi$. Here it arises as the only solution to a system of equations derived from scattering amplitudes.  Later on, we give an analytical derivation of the DBI Lagrangian from the soft limit.

\medskip
\noindent {\bf Case: $\rho=2$}
\medskip

The next case to consider is $\rho=2$. The inequality in Eq.~\ref{eq:interesting} imply that $\sigma\geq2$ to have a non-trivial case. Here we consider a generic action which has $2n-2$ derivatives on $n$ fields $\phi$. The theory must have at least one derivative per field, but the $n-2$ derivatives can be distributed in various ways among fields, so schematically
${\cal L}_{(2,1)} \sim \sum_{n=2}^\infty F_n(\partial^{2n-2}\phi^n)$,where $F_n$ denotes a collection of operators with free coefficients which have $2n-2$ derivatives on $n$ fields. We construct the ansatz for the amplitude for a given $n$ and impose the condition that $A_{n,1}=0$, so $A_n = {\cal O}(z^2)$. For $n=4$ there are two independent kinematical structures
\begin{equation}
A_4 = c_1(s_{12}^3+s_{23}^3+s_{31}^3) + c_2 (s_{12}s_{23}s_{31}),
\end{equation}
whose behavior is ${\cal O}(z^3)$ for arbitrary $c_1$ and $c_2$, as argued earlier. Going to higher points we find unique solutions for the constraints: $A_5={\cal O}(z^2)$, $A_6={\cal O}(z^3)$, $A_7={\cal O}(z^2)$ while for $n=8$ we get two solutions for $\sigma=2$ and one solution for $\sigma=3$. It is easy to see the amplitude is generated by the Lagrangian
\begin{equation}
{\cal L}_{(2,2)} = \lambda_4 {\cal O}_4 + \lambda_5 {\cal O}_5 ,
\end{equation}
where ${\cal O}_4\sim~\partial^6\phi^4$ and ${\cal O}_5\sim \partial^8\phi^5$ are four and five point interaction vertices.  Indeed, the derivative counting and structure is precisely that of the four and five point interaction vertices of the four dimensional Galileon theories studied in \cite{galbas}.  Famously, the Galileon Lagrangian exhibits a second order shift symmetry $\phi\to\phi +a+b_\mu x^\mu$ and has equations of motion that are second order in derivatives of $\phi$. The missing three point interaction can be eliminated via the Galileon duality (for a detailed discussion see \cite{Kampf:2014rka}), yielding just the four and point interactions, which we denote by Galileon$_{4,5}$.

Intriguingly, we have checked up to twelve particles that the amplitudes derived from ${\cal O}_4$ alone yield $A_n={\cal O}(z^3)$, which suggests an even simpler theory
\begin{equation}
{\cal L}_{(2,3)} = \lambda_4 {\cal O}_4,
\end{equation}
which we will refer to as Galileon$_4$.

\medskip
\noindent {\bf Case: $\rho>2$}
\medskip

We have done some partial analyses for $\rho=3$ and $\rho=4$ for $n=5$ and for $\rho=3$ for $n=6$ and indeed there are unique amplitudes there with non-trivial soft-limit behavior, {\it i.e.}~$A_5,A_6 = {\cal O}(z^3)$ for $\rho=3$ and $A_5={\cal O}(z^4)$ for $\rho=4$. It is very suggestive that these are exactly the theories found in \cite{shiftsym}, {\it i.e.}~theories with higher shift symmetries. 

\medskip
\noindent {\bf Case: $1<\rho=$ fractional}
\medskip

As discussed earlier, $\rho$ is a non-negative rational number. Restricting to derivatively coupled theories, $\rho\geq 1$, so we should consider all theories with $\rho=m/n$ for integers $m,n$ with $m\geq n$. For example for $\rho=3/2$ we have $\sigma\geq2$, and the schematic Lagrangian is
\begin{equation}
{\cal L}_{(\rho,\sigma)} \sim (\partial\phi)^2 + (\partial^8\phi^6) + (\partial^{14}\phi^{10}) + \dots
\end{equation}
For this case we have checked that ${\cal O}(z^2)$ soft behavior is impossible with $(\partial^8\phi^6)$, and first becomes possible with $(\partial^{14}\phi^{10})$. We have done this check for $\rho=\frac32, \frac43, \frac65, \frac85$ which rules out all theories with operators $n<8$ for $\sigma=2$. Further analysis is needed to generate more data and also get a better intuition how to prove that such theories do or do not exhibit interesting soft-limit behavior.

\section{Amplitudes from Equations of Motion}

Our analysis thus far does not prove the existence of theories, as such a claim would require an accounting of an infinite number of amplitudes.  Instead we have demonstrated that theories with certain $(\rho,\sigma)$ can or cannot exist. We can then finish the job by using the prescribed soft limits to construct the Lagrangians for these theories explicitly.   In what follows, we show how this can be done with the equations of motion.  The action for a massless scalar field is
\begin{equation}
S[\phi ] =  \int \mathrm{d}^{4}x\; {\cal L}(\phi,\partial\phi) = \int \mathrm{d}^{4}x\; \frac{1}{2}(\partial \phi)^2 + S_{\rm int}[\phi ],
\end{equation}
where $S_{\rm int}$ contains all non-linear interactions.  The equations of motion are
\begin{align}
\Box \phi(x) = \frac{\delta S_{\rm int}[\phi]}{\delta \phi(x)}.
\label{eq:EOM}
\end{align}
Equations of motion are identically satisfied when evaluated inside a time-ordered product.  Hence, Eq.~\ref{eq:EOM} is valid when sandwiched between in and out states like that define the on-shell scattering amplitude $A = \langle f | i\rangle$.  We thus obtain  $\langle f|{\delta S_{\rm int}[\phi]}/{\delta
 \phi (x)}|i\rangle =  \langle f| \Box
\phi (x)|i\rangle$.  Transforming to the Fourier conjugate field $\tilde \phi(p)$ where $p \equiv p_f - p_i$ and taking the on-shell limit, $p^2 \rightarrow 0$, we obtain the LSZ reduction formula for the amplitude with the scalar emitted with momentum $p$,
\begin{align}
A(p) &\equiv \langle f +\tilde \phi(p) | i\rangle 
=
\lim_{p^{2}\rightarrow 0}
i\langle f |\frac{\delta S_{\rm int}[\tilde\phi]}{\delta
\tilde \phi (p)}|i\rangle .
\label{eq:LSZ}
\end{align}
%
%
%
%
%
Here $A(p)$ satisfies Eq.~\ref{eq:soft} in the soft limit if and only if 
\begin{align}
\langle f|\frac{\delta S_{\rm int}[\phi ]}{\delta \phi (x)}
| i \rangle &= \partial_{\mu_1}\dots\partial_{\mu_\sigma}\langle 
f|K^{\mu_1\dots \mu_\sigma }(x)|i\rangle  ,
\label{condition1}
\end{align}
for some local operator $K^{\mu_1\ldots\mu_n}(x)$, in which case
\begin{align}
A(p) &= \lim_{p^{2}\rightarrow 0} i^{\sigma+1} p_{\mu_1}\dots p_{\mu_\sigma}\langle 
f|\tilde K^{\mu_1\dots \mu_\sigma }(p)|i\rangle  .
\end{align}
The assumption of locality is crucial: if the Fourier transformed operator $\tilde K^{\mu_1\ldots\mu_n}(p)$ is singular as $p$ goes to zero, then this will compensate for  the  $ p_{\mu_1}\dots p_{\mu_\sigma}$ factors in the numerator.   However, regularity of the operator at $p\rightarrow0$ is in principle violated if the theory has cubic vertices, in which case the soft limit can generate collinear singularities which produce inverse powers of $z$.  

The condition in Eq.~\ref{condition1} is satisfied provided
\begin{align}
\frac{\delta S_{\rm int}[\phi ]}{\delta \phi (x)} &= \partial_{\mu_1}\dots\partial_{\mu_\sigma}K^{\mu_1\dots \mu_\sigma }(x),
\label{eq:condition2}
\end{align}
on the support of any equations which hold when evaluated inside the time-ordered product, {\it i.e.}~when sandwiched between the in and out states.  A priori, Eq.~\ref{eq:condition2} can be true due to algebraic identities or conservation equations.  Let us classify each theory in turn.

\medskip
\noindent \textbf{Case:} $\left( \rho ,\sigma \right) =\left( 1,1\right) $
\medskip

These theories have exactly one derivative per field so Lagrangian is $%
\mathcal{L}\left( X\right) =X/2+\mathcal{L}_{\mathrm{int}}\left( X\right)
=\sum_{n}c_{n}X^{n}$ where $X=\left( \partial \phi \right) ^{2}$. The Noether current is 
\begin{equation}
J^{\mu }=\frac{\partial \mathcal{L}\left( X\right) }{\partial \left(
\partial _{\mu }\phi \right) }=2\mathcal{L}^{\prime }\left( X\right)
\partial ^{\mu }\phi ,
\end{equation}
and the variation of the action yields
\begin{equation}
\frac{\delta S_{\rm int}}{\delta \phi }=-\partial _{\mu }\frac{\partial \mathcal{%
L}_{\mathrm{int}}\left( X\right) }{\partial \left( \partial _{\mu }\phi
\right) }=\partial _{\mu }\left( \partial ^{\mu }\phi -J^{\mu }\right) ,
\label{eq:eqmot}
\end{equation}
which is of the form of Eq.~\ref{eq:condition2} with $\sigma =1$ for any $c_{n}$. 

\medskip
\noindent \textbf{Case:} $\left( \rho ,\sigma \right) =\left( 1,2\right) $
\medskip

Extending to an enhanced soft limit  $\sigma =2$ implies additional
constraints on $\mathcal{L}(X)$, so the $c_{n}$ are constrained. 
Plugging Eq.~\ref{eq:eqmot} into Eq.~\ref{condition1} for $\sigma=2$ implies that
\begin{equation}
\langle f|J^{\mu }|i\rangle =\partial _{\nu }\langle f|L^{\mu \nu }|i\rangle .
\label{eq:fJi}
\end{equation}
However, $J^{\mu }=\partial _{\nu }L^{\mu \nu }$  cannot be true
algebraically, simply because $L^{\mu \nu }$ involves $\partial \phi $, so $%
\partial _{\nu }L^{\mu \nu }$ involves $\partial \partial \phi $, which can
never match $J^{\mu }$, which only involves $\partial \phi $. Consequently, we need a supplemental equation that holds when evaluated between in and out states.  The natural candidate equation is conservation of the energy-momentum tensor, 
\begin{equation}
T^{\mu \nu }=2\mathcal{L}^{\prime }\left( X\right) \partial ^{\mu }\phi
\partial^{\nu }\phi -\eta ^{\mu \nu }\mathcal{L}\left( X\right).
\end{equation}
By derivative counting, there is a unique ansatz for $L^{\mu\nu}$ for which $\partial_\nu L^{\mu\nu}$ does not involve $\partial\partial \phi$ when evaluated between in and out states: $L^{\mu\nu} = g \phi T^{\mu \nu}$ for some constant $g$.  Plugging this into Eq.~\ref{eq:fJi} we obtain
\begin{equation}
\langle f|J^{\mu }|i\rangle =\partial _{\nu }\langle f|L^{\mu \nu }|i\rangle = g\langle f |T^{\mu \nu }\partial _{\nu }\phi |i\rangle,
\end{equation}
where we have used that $\partial_\nu T^{\mu\nu}=0$ when evaluated between in and out states.
This formula is automatically true if $J^{\mu }=gT^{\mu \nu
}\partial _{\nu }\phi $ applies algebraically, which implies the differential equation%
\begin{equation}
2\mathcal{L}^{\prime }\left( X\right) /g=2\mathcal{L}^{\prime }\left( X\right)
X-\mathcal{L}\left( X\right) ,
\end{equation}
whose solution is $\mathcal{L}\left( X\right) \propto \sqrt{1-gX}$, which is
precisely the DBI action for a single scalar field. We might wonder if there are some other
solutions which do not satisfy $J^{\mu }=gT^{\mu \nu }\partial _{\nu }\phi $
algebraically. However, our earlier scattering amplitudes analysis found that there existed only one or zero theories with a $\sigma=2$ soft limit, so the one and only solution is DBI.

\medskip
\noindent {\bf Case: $(\rho,\sigma) =(2,2)$}
\medskip

These theories are of the form ${\cal L} = \sum_n c_n \partial^{2(n-1)} \phi^n$.  As before, $\sigma=2$ implies the condition in Eq.~\ref{eq:fJi}.  As shown in \cite{Hinterbichler:2010xn}, this constraint is algebraically satisfied for the Noether current associated with the Galileon.  A subtlety in this case is that the Galileon theory has a cubic vertex, in which case $L^{\mu \nu}$ can in principle be not regular at $p\rightarrow0$.  However, as shown in \cite{Kampf:2014rka}, a field redefinition of the Galileon can be used to eliminate the cubic vertex, truncating down to Galileon$_{4,5}$.  Note that in an explicit evaluation of amplitudes shows the further restriction to only four point interaction vertices, Galileon$_4$, satisfies stronger soft-limit behavior, $\sigma=3$. We do not yet have a general proof for this.  

Finally, we note that for $(\rho,\sigma)=(1,2),(2,2)$, one can construct a fully non-perturbative derivation of ${\cal O}(z^2)$ scaling from 
symmetries without a priori knowledge of the Lagrangian. We present the full details in \cite{inprep}.

\section{Conclusion}

In this letter we have constructed scattering amplitudes for a massless scalar in order to derive and classify effective field theories. Enhanced soft limits largely fix the Lagrangians of the corresponding effective field theories, and we have presented derivations of DBI and the Galileon as examples.  This work is part of a more general program to construct and classify all possible effective field theories from their soft limits.  Intriguingly, the amplitudes approach allows for the possibility of discovering new effective field theories and symmetries. 

\smallskip

{\it Acknowledgment:} We thank Nima Arkani-Hamed for useful discussions and comments. C.C.~is supported by a Sloan Research Fellowship and a DOE Early Career Award under Grant No.~DE-SC0010255. J.T.~is supported in part by the David and Ellen Lee Postdoctoral Scholarship and by the Department of Energy under grant number DE-SC0011632.~K.K.~and J.N.~are supported by project PRVOUK P45 and grants LG 13031 and LH 14035.

\end{document}